\documentclass[12pt]{article}
\usepackage{amssymb}

\title{Exact solution of Riemann--Hilbert problem for
  a correlation function of the $XY$ spin chain}
\author{Yasuhiro Fujii}

\date{\textit{\small Department of Physics, Graduate School of Science,
  University of Tokyo,\\  Hongo 7-3-1, Bunkyo-ku, Tokyo 113-0033, Japan}}

\newcommand{\e}{\mbox{e}}
\renewcommand{\d}{\mbox{d}}

\renewcommand{\i}{\mbox{i}}
\newcommand{\ii}{\mbox{\scriptsize i}}
\newcommand{\Det}{\mbox{Det}}
\newcommand{\C}{{\mathbb C}}
\renewcommand{\epsilon}{\varepsilon}

\begin{document}
\maketitle

\begin{abstract}
  A correlation function of the $XY$ spin chain is studied
  at zero temperature.
  This is called the Emptiness Formation Probability (EFP)
  and is expressed by the Fredholm determinant
  in the thermodynamic limit.
  We formulate the associated Riemann--Hilbert problem
  and solve it exactly.
  The EFP is shown to decay in Gaussian.
\end{abstract}

In this letter we study the Riemann--Hilbert problem associated with
a correlation function of the $XY$ spin chain.
The Hamiltonian of the model is given by
\begin{equation}
  H_{XY} =
  \sum_{n=1}^N(\sigma_n^x\sigma_{n+1}^x+\sigma_n^y\sigma_{n+1}^y
  -h\sigma_n^z).
\end{equation}
Here $\sigma_n^x$, $\sigma_n^y$, $\sigma_n^z$ are the Pauli matrices
acting on the $n$-th site and $h$ indicates an external magnetic field.
Let $p_k=(1+\sigma_k^z)/2$.
Consider a correlation function,
\begin{equation}
  P_n = \langle p_1\cdots p_n\rangle,
\end{equation}
which describes the probability of finding a string of
$n$-adjacent parallel spins up on the ground state of the model
for a given value of the external magnetic field $h$.
This is called the Emptiness Formation Probability (EFP).
In the thermodynamic limit ($N\rightarrow\infty$),
at zero temperature,
the EFP is known to be expressed by the Fredholm determinant
as follows \cite{CIKT}:
\begin{equation}
  P_n = \Det(1-K_n),
\end{equation}
where the kernel $K_n$ is of the form,
\begin{equation}
  \label{kn}
  K_n(z,w) = \frac{f^T(z)g(w)}{z-w},
\end{equation}
with two-component vectors
\begin{eqnarray}
  \label{vec1}
  f(z) &=& \left(
    \begin{array}{c}
      1 \\ z^n
    \end{array}
  \right),
  \\
  \label{vec2}
  g(z) &=& \frac{1}{2\pi\i}\left(
    \begin{array}{c}
      -1 \\ z^{-n}
    \end{array}
  \right).
\end{eqnarray}
We remark that the kernel $K_n$ has no singularities because
\begin{equation}
  \label{fg}
  f^T(z)g(z) = 0.
\end{equation}

This kind of Fredholm determinant can be computed 
by means of a classical inverse scattering problem,
the so-called Riemann--Hilbert problem.
By a heuristic method
the \textit{asymptotic} solution has been derived \cite{DIZ}.
In this letter we propose a systematic approach
based on the analytic properties
of the vectors (\ref{vec1})--(\ref{vec2})
and solve the Riemann--Hilbert problem \textit{exactly}.
The EFP is shown to decay in Gaussian.

The Riemann--Hilbert problem related to the EFP
is formulated as follows.
Define a two-component vector by
\begin{equation}
  F(z) = (1-K_n)^{-1}f(z).
\end{equation}
Let $L$ be an open contour;
\begin{equation}
  L = \{z:|z|=1,\phi<\mbox{arg}z<2\pi-\phi\},
\end{equation}
with a parameter $\pi/2<\phi<\pi$,
which is connected with the external magnetic field $h$ through
\begin{equation}
  \label{hf}
  \cosh 2\Lambda = \frac{2}{h},
  \qquad
  \e^{\ii\phi} = -\i\frac{\e^{-2\Lambda}-\i}{\e^{-2\Lambda}+\i}.
\end{equation}
Here it is assumed that $0<h<2$.
The contour $L$ thus runs on a part of the unit circle anti-clockwise.
Consider a $2\times 2$ matrix function;
\begin{equation}
  \label{Y}
  Y(z) = I -\int_L\frac{\d w}{w-z}F(w)g^T(w),
\end{equation}
where $I$ is the $2\times 2$ unit matrix.
Let $Y_+(z)$ ($Y_-(z)$) be the boundary value of $Y(z)$
onto the contour $L$ from the left (the right) of the positive direction
of the contour $L$.
For $z\in L$ the Plemelj formulae yield
\begin{equation}
  \label{y-y}
  Y_+(z)-Y_-(z) = -2\pi\i F(z)g^T(z),
\end{equation}
which implies (recall eq.~(\ref{fg})),
\begin{equation}
  Y_+(z)f(z)=Y_-(z)f(z).
\end{equation}
Using the expression of the kernel (\ref{kn}) we have
\begin{eqnarray}
  \label{yF}
  Y(z)f(z) &=&
  f(z)-\int_L\frac{\d w}{w-z}F(w)g^T(w)f(z)
  \nonumber \\ &=&
  f(z)+\int_L\d w K_n(z,w)F(w)
  \nonumber \\ &=&
  F(z).
\end{eqnarray}
It is thus made clear that $F(z)$ is an entire function
and is given by
\begin{equation}
  \label{Fy}
  F(z) = Y(z)f(z).
\end{equation}
Applying this relation to eq.~(\ref{y-y}) we obtain
\begin{equation}
  Y_+(z) = Y_-(z)(I-2\pi\i f(z)g^T(z)).
\end{equation}
Supplemented by analytic properties of the Cauchy-type integral,
this equation shows that $Y(z)$ is the solution
of the following $2\times 2$ matrix Riemann--Hilbert problem: \\
(a) $Y(z)$ is holomorphic for any $z\in\C\backslash L$. \\
(b) $Y(z)\rightarrow I$ as $z\rightarrow\infty$. \\
(c) $Y_+(z)=Y_-(z)H(z)$ for $z\in L$,
where the jump matrix $H(z)$ is expressed by
\begin{equation}
  H(z) = \left(
    \begin{array}{cc}
      2 & -z^{-n} \\ z^n & 0
    \end{array}
  \right).
\end{equation}

A correlation function EFP is related to
the Riemann--Hilbert problem as follows.
In terms of the solution of the Riemann--Hilbert problem (a)--(b)
the EFP can be computed
according to the following recursion relation:
\begin{equation}
  \frac{P_{n+1}}{P_n} = Y_{22}(0).
\end{equation}
The proof is straightforward.
By the explicit form of the kernel (\ref{kn})
it follows that
\begin{equation}
  K_{n+1}(z,w) = K_n(z,w)+E_n(z,w),
\end{equation}
where
\begin{equation}
  E_n(z,w) = \frac{1}{2\pi\i}z^n w^{-n-1} =
  \frac{1}{2\pi\i}f_2(z)w^{-n-1}.
\end{equation}
Hence
\begin{eqnarray}
  P_{n+1} &=& P_n\Det(1-(1-K_n)^{-1}E_n)
  \nonumber \\ &=&
  P_n\left(1-\frac{1}{2\pi\i}\int_L\d w F_2(w)w^{-n-1}\right).
\end{eqnarray}
On the other hand, from the definition (\ref{Y}),
\begin{eqnarray}
  Y_{22}(0) &=&
  1-\int_L\frac{\d w}{w}F_2(w)g_2^T(w)
  \nonumber \\ &=&
  1-\frac{1}{2\pi\i}\int_L\d w F_2(w)w^{-n-1}.
\end{eqnarray}
The proof is complete.

In general, there is no method
to solve the (matrix) Riemann--Hilbert problem.
In the only case that the jump matrix is triangular
the problem has been solved exactly \cite{AF}.
In order to solve our problem (a)--(c)
we pay attention to the fact that $F(z)$ is an entire function.
Because of eq.~(\ref{Fy}) and the condition (a)--(b),
$F(z)$ must be a vector consisting of polynomials as follows:
\begin{equation}
  F(z) = \left(
    \begin{array}{c}
      Q_n(z) \\ P_n(z)
    \end{array}
  \right),
\end{equation}
where
\begin{eqnarray}
  Q_n(z) &=& \sum_{k=0}^{n-1}q_k z^k,
  \\
  P_n(z) &=& z^n+\sum_{k=0}^{n-1}p_k z^k.
\end{eqnarray}
By virtue of this expression
the solution of the Riemann--Hilbert problem (a)--(c)
is expressed in terms of polynomials $Q_n(z)$, $P_n(z)$ as follows:
\begin{eqnarray}
  \label{y11}
  Y_{11}(z) &=& 1+\frac{1}{2\pi\i}\int_L
  \frac{\d w}{w-z}Q_n(w), \\
  Y_{12}(z) &=& -\frac{1}{2\pi\i}\int_L
  \frac{\d w}{w-z}w^{-n}Q_n(w), \\
  Y_{21}(z) &=& \frac{1}{2\pi\i}\int_L
  \frac{\d w}{w-z}P_n(w), \\
  \label{y22}
  Y_{22}(z) &=& 1-\frac{1}{2\pi\i}\int_L
  \frac{\d w}{w-z}w^{-n}P_n(w).
\end{eqnarray}
This solution is unique:
the scalar function $\det Y(z)$ has no jump
because $\det H(z)\equiv 1$.
In the limit $z\rightarrow\infty$
the expressions of the solution (\ref{y11})--(\ref{y22})
yields $\det Y(z)\equiv 1$
and thus the inverse of $Y(z)$ exists for any $z\in\C$,
by Liouville's theorem.
Let $\widetilde{Y}(z)$ be another solution
of the Riemann--Hilbert problem (a)--(c).
Then $\widetilde{Y}(z)Y^{-1}(z)$ has no jump on the contour $L$ and
the condition at infinity leads to $\widetilde{Y}(z)Y^{-1}(z)\equiv 1$,
that is, $\widetilde{Y}(z)\equiv Y(z)$.

Let us determine the explicit forms of polynomials $Q_n(z)$, $P_n(z)$.
Expanding the right hand side of eq.~(\ref{Fy})
with respect to $z$ we have
\begin{eqnarray}
  \left(
    \begin{array}{c}
      q_k \\ p_k
    \end{array}
  \right) =
  \left(
    \begin{array}{c}
      \delta_{k0} \\ 0
    \end{array}
  \right)
  -\sum_{l=0}^n\frac{\sin\phi(k-l)}{\pi(k-l)}\left(
    \begin{array}{c}
      q_l \\ p_l
    \end{array}
  \right),
  \nonumber \\
\end{eqnarray}
where $k=0,\ldots,n-1$ and $q_n=0$, $p_n=1$.
In terms of an $(n+1)\times(n+1)$ matrix $S$;
\begin{eqnarray}
  S_{ij} &=& -\frac{\sin\phi(i-j)}{\pi(i-j)},
  \\
  S_{nj} &=& 0,
  \qquad
  (i=0,\ldots,n-1,\quad j=0,\ldots,n)
  \nonumber
\end{eqnarray}
the coefficients of polynomials $Q_n(z)$, $P_n(z)$
can be expressed by
\begin{eqnarray}
  \label{qp}
  q_k &=& (1-S)_{k0}^{-1},
  \\
  p_k &=& (1-S)_{kn}^{-1}.
  \qquad
  (k=0,\ldots,n)
  \nonumber
\end{eqnarray}
We thus obtain the exact and unique solution
of the Riemann--Hilbert problem (a)--(c)
associated with a correlation function EFP
by virtue of our systematic approach based on
the analytic property of the vector $F(z)$.
The solution is given by eqs.~(\ref{y11})--(\ref{y22})
and the coefficients of polynomials included in them
are determined by eqs.~(\ref{qp}).

As an example we consider a limiting case that
a value of the external magnetic field tends to $2$
(strong magnetic field limit).
By the relations (\ref{hf})
the boundary parameter $\phi$ is then put into
$\pi(1-\epsilon)$ and the solution $Y(z)$ can be expanded with $\epsilon$.
Up to the order $\epsilon^3$ the matrix $S$ is expanded as
\begin{eqnarray}
  S_{ij} &=& -\delta_{ij}+(-1)^{i-j}\epsilon+O(\epsilon^3),
  \\
  S_{nj} &=& 0,
  \qquad
  (i=0,\ldots,n-1,\quad j=0,\ldots,n)
  \nonumber
\end{eqnarray}
and the inverse of $1-S$ is computed as
\begin{eqnarray}
  (1-S)_{ij}^{-1} &=&
  \delta_{ij}-\frac{(-1)^{i-j}}{2-\delta_{jn}}\delta_{i\ne n}
  \left(\delta_{ij}-\frac{\epsilon}{2}-\frac{n\epsilon^2}{4}\right)
  \nonumber \\ &&
  +O(\epsilon^3).
  \qquad
  (i,j=0,\ldots,n)
\end{eqnarray}
Here $\delta_{i\ne n}$ yields $1$ unless $i=n$.
We thus have
\begin{eqnarray}
  q_k &=&
  \frac{\delta_{k0}}{2}+\frac{(-1)^k}{4}
  \delta_{k\ne n}\left(\epsilon+\frac{n\epsilon^2}{2}\right)
  +O(\epsilon^3), \\
  p_k &=& \delta_{kn}+\frac{(-1)^{n-k}}{2}\delta_{k\ne n}
  \left(\epsilon+\frac{n\epsilon^2}{2}\right)
  +O(\epsilon^3).
  \nonumber \\ && \qquad\qquad\qquad
  (k=0,\ldots,n)
\end{eqnarray}
In order to derive the asymptotic form of the EFP
we calculate $Y_{22}(0)$ up to the order $\epsilon^3$:
\begin{eqnarray}
  Y_{22}(0) &=&
  1-\frac{1}{2\pi\i}\int_L\frac{\d w}{w}
  \sum_{k=0}^n p_k w^{k-n}
  \nonumber \\ &=&
  2-\epsilon-\frac{n\epsilon^2}{2}+O(\epsilon^3).
\end{eqnarray}
Taking this logarithm we obtain the following recursion relation:
\begin{eqnarray}
  \log P_{n+1}-\log P_n =
  \log 2-\frac{\epsilon}{2}-\frac{\epsilon^2}{8}(2n+1)+O(\epsilon^3),
  \nonumber \\
\end{eqnarray}
which implies
\begin{equation}
  \label{result}
  \log P_n =
  -\frac{\epsilon^2}{8}n^2+\left(\log2-\frac{\epsilon}{2}\right)n
  +O(\epsilon^3).
\end{equation}
The EFP is thus shown to decay in Gaussian:
$P_n\sim\alpha^{-n^2}$ with some value $\alpha>1$,
up to the order $\epsilon^3$.

The asymptotic form of the EFP has been computed
by a heuristic approach to Riemann--Hilbert problem
for sufficiently large $n$ \cite{DIZ}.
The result is
\begin{equation}
  \log P_n = n^2\log\sin\frac{\phi}{2}+O(n),
\end{equation}
and is expanded as
\begin{equation}
  \log P_n = -\frac{\epsilon^2}{8}n^2+O(\epsilon^3,n),
\end{equation}
where we set $\phi=\pi-\epsilon$.
This coincides with our result (\ref{result}),
up to the order $n$.

In this letter we have considered the Riemann--Hilbert problem
associated with the EFP\@.
Based on the analytic properties of the vectors
the Riemann--Hilbert problem has been shown to be solved exactly.
It has made clear that the EFP decays in Gaussian.

Recently, the Riemann--Hilbert problem related to
the generating functional of correlation functions
of the $XXZ$ spin chain has been established \cite{FW1,FW2}.
The jump matrix of this problem is expressed by a $4\times 4$ matrix.
In the same way as this letter
the Riemann--Hilbert problem for the $XXZ$ spin chain
can be solved exactly.
In a forthcoming publication
we will solve this Riemann--Hilbert problem
and analyze the asymptotic forms of two-point correlation functions
of the $XXZ$ spin chain.

\section*{Acknowledgements}
The author is grateful for a research fellowship
of the Japan Society for the Promotion of Science (JPSJ)
for young scientists.


\begin{thebibliography}{99}
\bibitem{CIKT} F. Colomo, A. G. Izergin, V. E. Korepin
  and V. Tognetti:
  \textit{Phys. Lett.} A {\bf 169} (1992) 243--247.
\bibitem{DIZ} P. A. Deift, A. R. Its and X. Zhou:
  \textit{Ann. Math.} {\bf 146} (1997) 149--235.
\bibitem{AF} M. J. Ablowitz and A. S. Fokas:
  \textit{``Complex Variables''}
  (Cambridge University Press, 1997, Cambridge).
\bibitem{FW1} Y. Fujii and M. Wadati:
  \textit{J. Phys.} A {\bf 33} (2000) 1351--1361.
\bibitem{FW2} Y. Fujii and M. Wadati:
  \textit{J. Phys.} A {\bf 33} (2000) 6497--6504.
\end{thebibliography}
\end{document}